# Conceptual Problems Related to Time and Mass in Quantum Theory

Daniel M. Greenberger
*Dept. of Physics, CCNY, New York, NY, 10031,USA.*
*greenbgr@sci.ccny.cuny.edu*

## Introduction

There are serious problems in quantum theory dealing with time order of events and trajectories of particles. But here, we shall concern ourselves with one very basic issue, concerning the meaning of proper time within quantum mechanics, and we shall see how this relates to the concept of mass. It is generally assumed that the concept of proper time can just be taken over from classical physics, and used straight-forwardly in quantum mechanics, and, so far as we know, very little thought has been given to just how large a divide exists between the classical concept of proper time, and the quantum-mechanical one. Consider as an example, that a particle propagating along meets and becomes trapped in a cavity, as in Fig. (1). In this case, representing the particle by a wave packet, one can write

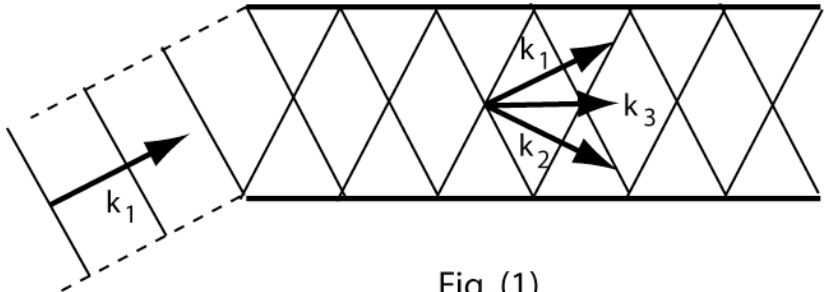

Fig. (1)

Fig. (1)- Proper Time Conflict. If a wave impinges on a cavity, standing waves will be set up inside. The waves inside are a superposition of two waves going with velocity $v$, while the composite wave moves with velocity $v \cos \theta$. Thus the proper time for the composite wave is greater than that of its components.

$$\psi(x,t) = \int a(\mathbf{k}_1) e^{i(k_1 \cdot r - \omega t)} d\mathbf{k}_1, \tag{1}$$

where $k$ is centered about some value $k_1$. Inside the cavity, due to internal reflections, there will be a transverse standing wave, the reflected part centered about $k_2$.

The total wave packet will then propagate in the direction

$$\tfrac{1}{2}(\mathbf{k}_1 + \mathbf{k}_2) = \mathbf{k}_3,$$
$$|\mathbf{k}_3| = |\mathbf{k}_1| \cos\theta, \quad |\mathbf{k}_1| = |\mathbf{k}_2|, \tag{2}$$

where $\theta$ is the angle between $k_1$ and the axis of the cavity.

The problem lies in the fact that the wave traveling along the axis of the cylinder has velocity $v_1 \cos \theta$, and so $\tau_3 = \sqrt{1 - v^2 \cos^2 \theta}\, t$, while it is a superposition of two traveling waves, each of which has a proper time of $\tau_1 = \sqrt{1 - v^2}\, t$. This is a clear ambiguity in the theory. One is tempted to resolve it with the quantum mechanical resolution that each of the component waves is an eigenstate of the momentum, and so the time $\tau_1$ is the correct proper time to use. But this conflicts with the relativistic velocity resolution. (In quantum theory, momentum trumps velocity!). But this should be resolved by experiment.

Another related problem, that is more enigmatic, concerns a particle inside a special interferometer, as in Fig. (2). Here, in one branch of the interferometer, the particle gets decelerated in an electric field, and then re-accelerated, so the amplitudes in each branch represent particles hitting the last beam splitter at the same speed. They also arrive at the same time, as the path of the other branch is increased to guarantee this. Thus the two branches interfere, and the result is counted at the detector D, where the two amplitudes have been recombined into the same particle, as in a usual beam splitter. Now the question is, what is the proper time of this particle, as the two branches have undergone different proper times. Furthermore, if the particle is unstable, when will it decay? Will the two amplitudes interfere, affecting the decay time? So far as we know, issues like this have never been decided experimentally.

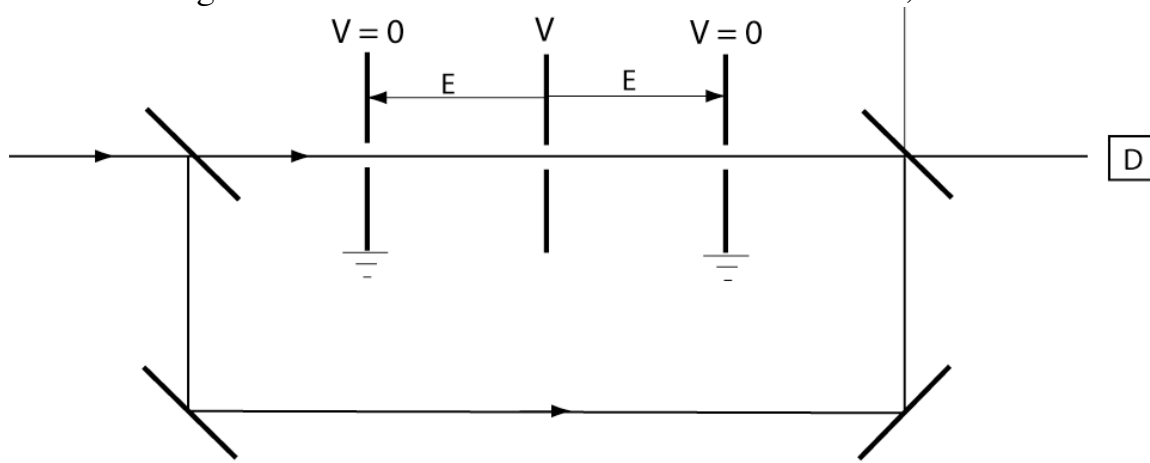

Fig. (2)

Fig. (2)- The Proper Time of a Superposition. Here a particle is split by an interferometer, and the two amplitudes take different proper times before recombining, which leads to questions about the subsequent behavior of the particle.

A different, but somewhat related, issue is that it has been shown that if one goes to the non-relativistic limit, there are residues of the proper time that show up in this limit, even though the concept of proper time is not defined there. An example of this is the Galilean transformation, which is a non-relativistic boost to a system moving with velocity $v$. In this system, there is a famous extra phase factor that shows up, with major consequences. Because of this, there is a theorem, due to Bargmann [1], that says that one cannot make a coherent superposition of two different mass states of a particle, non-relativistically.

This theorem follows directly from this phase factor, which is mass dependent. By considering a series of transformations to a set of coordinate systems moving at different velocities, and ending up back at the original system, one finds that one has induced an accumulated phase change for each wave function in the superposition, and since the phase is mass-dependent, each mass component will have accumulated a different phase. One could detect this phase difference by an interference experiment, if it exists. However, this makes no sense non-relativistically, since one ends up in the same state that one started from, and so this phase difference cannot be real. One eliminates it by the super-selection rule that one cannot superimpose states of different masses.

But this theorem is very troubling, as relativistically, since mass is a form of energy, one can certainly superimpose states of different mass, as in $K^0$ decay. So how can a greater symmetry, namely Lorentz Invariance, in the limit of the less symmetrical non-relativistic theory, produce an extra selection rule? Usually, it is the other way round, where the extra symmetry produces extra restrictions on the states. The answer is found in the nature of the extra phase that comes in non-relativistically. It turns out that this phase is nothing more than the twin paradox in the non-relativistic limit [2], and is proportional to $e^{im(t-\tau)/\hbar}$, and appears differently for the two mass states. (Throughout, we shall set $c = 1$.) However, relativistically, the effect is very real, and there is a difference between the case of having made the excursion and not having made it, namely the twin paradox. So this is a physically meaningful phase effect, and it is experimentally realizable. But in the non-relativistic case, it is embarrassing, because here is a real effect and no interpretation for it, since the notion of proper time is not recognized. So one rules it out by fiat. But it should be experimentally observable, even in a non-relativistic experiment. So the Bargmann solution is untenable.

Thus, proper time leaves a residue in certain non-relativistic problems, and this should be recognized. And the same is true of rest mass, which also shows up here. The concepts of rest mass and proper time are omitted in non-relativistic theory. Classically, one can get away with this, but they leave residues in quantum theory and we believe that it is a mistake to ignore them, and that they should have very important roles to play. We shall further explore this thought below.

As for the mass, there is an immediate clue as to the role it should play in quantum theory. The role it actually plays in present theories is as an independent parameter. It plays no dynamical role in the theory, but is placed into the theory by hand, each independent field having its own pre-determined mass. If two particles interact, the energy of the system can dynamically change through its binding energy, by way of the interaction, but there is no dynamical way for the mass to correspondingly change to keep pace with it. Again, one can consistently put this in by hand. (This should not be confused with the fact that the rest mass itself can be controlled by internal symmetries, like SU(3).)

The clue I referred to is that for a free particle, there is no mass. One free particle is like any other. It is only through a non-gravitational interaction with the environment that a particle acquires its mass. For example, if a particle decays into two, of known rest masses $m_1$ and $m_2$, one does not know the masses of either one until it interacts with something. For example, if one of them passes through a slit, it acquires a deBroglie wavelength, which is mass-dependent. It is only then that one knows which type of particle passed through the slit. If it passes instead through an external gravitational field, the mass drops out by the equivalence principle, and one still doesn't know which type of particle it is.

In the usual way in which we treat quantum theory, a free particle propagating along already has a mass, and a deBroglie wavelength. But this is unmeasured, and untested, and violates the basic tenets of quantum theory. Once the particle interacts with something, we can detect its mass, but we usually say it has always had this mass, and we won't make any mistakes in doing this, since it would be hard to tell the difference. But actually, this is EPR type thinking, not quantum thinking. We credit the particle with a mass state from the time it was produced, although we say it doesn't have an actual spin state until we measure it. Clearly there is an inconsistency here and there is probably a Bell-type theorem that could tell the difference.

## The Mass and Proper Time as Operators

What is the way out? The mass should be treated as a quantum-mechanical operator. Until the particle interacts it is in a superposition of different mass states, and the interaction puts it into one specific eigenstate of the mass, just as with any other dynamical variable. So the mass should enter the theory as a dynamical variable.

Similarly with the proper time. A particle does not acquire a proper time until something like a trajectory, or set of trajectories, can be determined through measurement. There is a further property that a classical dynamical variable has which we must explore here for the case of proper time. For a particle, the position *x* alone is not sufficient to determine its status as a dynamical variable. One can always independently set the initial conditions, and so one needs both the initial position *and* the initial velocity. Then the equations of motion are sufficient to determine the further propagation of the particle in time. One can replace this with the initial position and the initial momentum.

When one looks at this for proper time, there is a problem within special relativity. If the proper time, $\tau$, which means the time as read by a clock located in the center of mass system of the particle (or relativistically, the barycentric system), is thought of as an independent dynamical variable, it is clear that one can always independently reset the clock. But within special relativity, one is not free to set $\dot{\tau}$, the equivalent of the velocity $\dot{x}$, as $\dot{\tau}$ is determined by geometry as $d\tau = \sqrt{1-v^2}\,dt$. However, this restriction is lifted in general relativity. In, say, a weak gravitational field, $g_{00}$ is determined by $g_{00} = (1+2\varphi)$, where $\varphi$ is the gravitational potential, and $d\tau = \sqrt{g_{\mu\nu}dx^{\mu}dx^{\nu}} = \sqrt{(1+2\varphi-v^2)}\,dt \xrightarrow[NR]{} (1-v^2/2+\varphi)dt$. So, e.g., one can place the entire system inside a thin spherical gravitational shell and although there will be no force, there will be a gravitational potential, which will reset the rate at which the proper time clock runs. Thus one is free to reset both $\tau$ and $\dot{\tau}$ independently, and thus the same conditions are met as one usually sees in classical physics. The momentum conjugate to the proper time will be the mass, and so, equivalently, one can have both independent proper times and masses.

Another interesting thing that happens for a free particle is that when the proper time is considered as a dynamical variable, the usual Hamiltonian for a free particle also yields the correct dynamical equation for the proper time. Normally, one has *H=H(x, p)*, but we have now extended it to
$$H = H(x,p;\tau,m), \tag{3}$$
and the proper time will be determined by dynamical interactions, rather than merely geometry. The normal equations of motion in *x* and *p* will be extended by two others in $\tau$ and *m*. The equations of motion become
$$\frac{\partial H}{\partial p} = \dot{x}, \quad -\frac{\partial H}{\partial x} = \dot{p};$$
$$\frac{\partial H}{\partial m} = \dot{\tau}, \quad -\frac{\partial H}{\partial \tau} = \dot{m}. \tag{4}$$
Then just as when one has a potential that depends on *x*, it leads to a force that changes the momentum, now if one has a potential that directly depends on $\tau$, it will lead to a "force" that changes the mass. So one now has a classical theory of decaying particles, something that one cannot do within conventional mechanics [3]. As an example of how natural an extension of the conventional theory this formalism is, consider the usual relativistic Hamiltonian for a free particle,
$$H = \sqrt{p^2 + m^2}, \tag{5}$$
which is symmetric in *p* and *m*. The standard equation of motion is

$$\dot{x} = v = \frac{\partial H}{\partial p} = \frac{p}{\sqrt{p^2 + m^2}},$$

$$p = \frac{mv}{\sqrt{1-v^2}}.$$

(6)

The second line of eq. (6) comes from inverting the first line to solve for $p$ in terms of $v$.

One also has now a second set of equations,

$$\dot{\tau} = \frac{\partial H}{\partial m} = \frac{m}{\sqrt{p^2 + m^2}} = \sqrt{1-v^2}.$$

(7)

So this equation, which was previously part of the geometry of space-time, has now become a dynamical equation of motion. The other two equations are simple here because there is no potential present,

$$\dot{p} = 0, \quad \dot{m} = 0.$$

(8)

However, one could add an external potential, $\varphi(\tau)$, and this would produce a decay in the mass.

One should note at this point that the mass is *not* the rest mass $m_0$. The mass is the energy in the rest system of the particle. Of course, in the above simple Hamiltonian, the mass is $m_0$. But in general, it must include, for example, the binding energies of the particles. The simplest case is what happens if one has two free particles. In that case, one has

$$H = \sqrt{p_1^2 + m_1^2} + \sqrt{p_2^2 + m_2^2}.$$

(9)

Call $p_1 + p_2 = P$, $E_1 + E_2 = E$. Then one can make a Lorentz Transformation into the Barycentric system, moving with velocity V such that its total momentum is 0,

$$P' = \gamma_V(P - VE), \quad E' = \gamma_V(E - VP).$$

(10)

Then, $V$ will be defined by $P' = 0$, so that $V = P/E$, and

$$E' \equiv M = \gamma_V E(1 - V^2),$$

$$E = \frac{M}{\sqrt{1-V^2}} = \sqrt{P^2 + M^2}.$$

(11)

This yields

$$E = E_1 + E_2 = \sqrt{(p_1 + p_2)^2 + M^2},$$

$$M = \sqrt{m_1^2 + m_2^2 + 2(E_1 E_2 - p_1 p_2)} \xrightarrow{NR} m_1 + m_2 + \frac{p^2}{2\mu},$$

(12)

where $\mu = \frac{m_1 m_2}{m_1 + m_2}$, and $p = \mu(v_1 - v_2)$.

So the mass includes the relative energy in the center of mass system, and if there were a potential present, it would include the complete binding energy, and it leaves out the energy of the center of mass, $P^2/2M$, which is just the rest energy

Lorentz transformed into the moving system, which does not change the mass of the system. In general, we will take the mass of the system to be the mass in the rest frame (barycentric frame, where $P' = 0$,) of the system. The rest energies of the individual particles, $m_1$ and $m_2$, we shall call the nominal masses of the particles. When particles interact, they no longer represent the true mass.

## The Uncertainty Relation Between the Mass and the Proper Time

It is a necessary consequence of considering the mass and proper time of a particle to be independent dynamical variables that they must obey an uncertainty relation. And one can show from many examples that there this is indeed the case and that an uncertainty principle exists, between the mass of a system and the proper time, that is, the time that would be read by a clock moving in the rest frame of the system. In other words, if one measures the mass of the system, the uncertainty in the result is directly related to the uncertainty in the proper time on the barycentric clock by an uncertainty principle, even if the time on a clock in the lab is known accurately. Many examples have been worked out, both in a gravitational field, and in a non-gravitational field, and we shall merely give an example of each.

First, imagine trying to measure the mass of a light particle by gravitationally scattering it off of a much heavier particle. Even though the mass of the lighter particle drops out of trajectory measurements, one can still determine its mass by getting its momentum involved. In scattering, the light particle will pick up a transverse momentum $p_x \sim FT$, where $T$ is the time it is in the strongly interactive region (see Fig. (3)), which will be of the order of $b/v$, where $b$ is the impact parameter in the figure. This gives $p_x \sim (GMm/b^2)T$. The uncertainty in the measurement will be $\delta m \sim \dfrac{b^2 \delta p_x}{GMT}$. However during the measurement, even if the time were known perfectly in the laboratory, while the particle is in the neighborhood of the heavy particle, its clock rate will be slowed by the presence of the gravity field of the heavy particle, and this effect will be of the order of $\tau \sim \varphi t$, or $\tau \sim (GM/b)T$. This will be uncertain to the order $\delta\tau \sim (GM/b^2)\delta b T$, and this gives

$$\delta m \delta\tau \sim \frac{b^2 \delta p_x}{GMT} \frac{GM \delta b T}{b^2} \sim \delta p_x \delta b \sim \hbar. \tag{11}$$

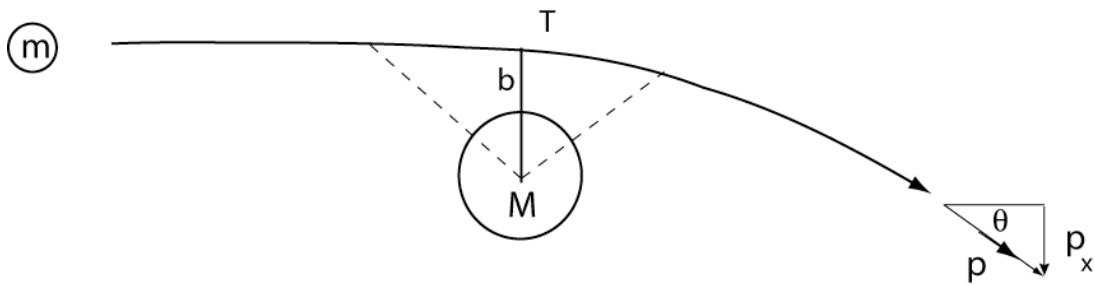

Fig. (3)

Fig. (3)- Gravitational Measurement of the Mass. The deflection of the lighter particle will determine its momentum and mass, while the red shift will determine its proper time. Together they give an uncertainty principle between the mass and proper time.

So the gravitational red shift affects the proper clock and the clock is uncertain to the extent that the position is uncertain, while the mass measurement is uncertain to the extent that the momentum is. The result is an overall uncertainty controlled by known uncertainty principles. By the same token, the famous Einstein-Bohr example of weighing a box of photons is also better described in terms of an uncertainty in mass and proper time, as there is an alarm clock sitting in the box that determines when the slit letting the photons out is to be opened. This clock is unattached to the laboratory, and is a true proper time clock for the box of photons.

As a non-gravitational example, imagine measuring the mass of a particle in a mass spectrometer, where there is a magnetic field perpendicular to the plane of the paper, as in Fig. (4), which bends the particle in a circle, the radius of which determines the mass of the particle. The force on the particle is

$$\begin{aligned} evB &= mv^2/R, \quad m = \frac{eBR}{v}, \\ \delta x &= 2\delta R \sim a, \quad \delta m \sim \frac{eB\delta R}{v} \sim \frac{eBa}{2}, \end{aligned} \tag{12}$$

where $a$ is the width of the entrance slit for the particle, which we take to be the biggest source of error in $R$. Because of the finite entrance slit, there will be diffraction by the particle, and the particles will pick up an uncertain $x$-momentum, and $\delta p_x \sim p\, \delta\theta$. This will cause the particle to travel a shorter time to the screen, by $\delta s \sim R\, \delta\theta$, taking a shorter time $\delta t \sim 2\delta s/v \sim 2R\delta\theta/v$. In this case, the proper time is related to the lab time not by the red shift, but rather by the special relativistic effect, which in the non-relativistic limit becomes $\delta\tau \sim (v^2/2)\, \delta t \sim vR\, \delta\theta$ $\sim vR\, \delta p_x/p$. But from eq. (12), $p/R = eB$, so $\delta\tau \sim v\delta p_x/eB$, and

$$\delta m\, \delta\tau \sim \frac{eBa}{2}\frac{v\delta p_x}{eB} \sim \hbar/2. \tag{13}$$

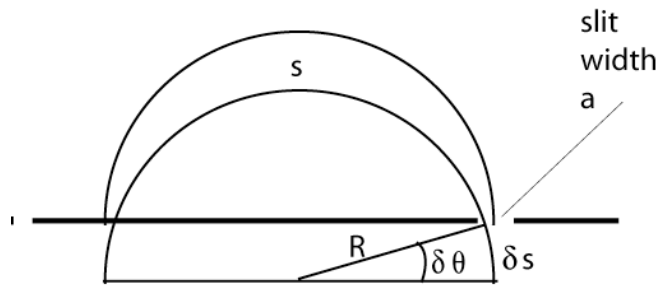

Fig. (4)

Fig. (4)- <u>Non-Gravitational Mass Measurement.</u> A particle enters a mass spectrograph, and the mass is determined by the radius R. But the slit causes diffraction, which will not only give a momentum uncertainty, but a proper time one as well.

So for both gravitational forces and non-gravitational ones, the uncertainty principle appears. This is one more indication of the necessity of treating the mass and proper time as operators.

There is an argument due to Pauli that says that the energy and lab time cannot be treated as operators, because the time, like the momentum, is unbounded. The momentum acts like a displacement operator for position, and this shows that the position must be unbounded. The same argument would show that the energy would be unbounded. But in our case, the situation is more complicated, because the proper time is not really unbounded, but has an upper limit given by the lab time.

So there are many real, unresolved issues concerning time, the perception of time, and the meaning of proper time, that arise in quantum theory. Connected to this, there are many problems connected with the concept of mass. It would certainly be a worthwhile project to sort them out, or at least make progress in that direction.